\documentclass[twocolumn,aps,prb,reprint,showpacs,amsmath,amssymb]{revtex4-1}
\usepackage{epsfig}
\usepackage{graphicx}% Include figure files
\usepackage{dcolumn}% Align table columns on decimal point
\usepackage{bm}% bold math
\usepackage[colorlinks=true,dvipdfm]{hyperref}

\usepackage{bookmark}

\begin{document}
\title{Scaling in driven dynamics starting in the vicinity of a quantum critical point}
\author{Shuai Yin}
\author{Chung-Yu Lo}
\author{Pochung Chen}
\affiliation{Department of physics, National Tsing Hua university, Hsinchu 30013, Taiwan}
\date{\today}

\begin{abstract}
We study the driven critical dynamics with an equilibrium initial state near a quantum critical point. In contrast to the original Kibble-Zurek mechanism, which describes the driven dynamics starting from an adiabatic stage that is far from the critical point, the  initial adiabacity is broken in this scenario. As a result, the scaling behavior cannot be described by the original Kibble-Zurek scaling. In this work we propose a scaling theory, which includes the initial parameters as additional scaling variables, to characterize the scaling behavior. In particular, this scaling theory can be used to describe the driven scaling behavior starting from a finite-temperature equilibrium state near a quantum critical point. We numerically confirm the scaling theory by simulating the real-time dynamics of the one-dimensional quantum Ising model at both zero and finite temperatures.
\end{abstract}

\maketitle

\section{\label{intro}Introduction}
\begin{figure}
  \centering
   \includegraphics[bb= 0 0 162 110, clip, scale=1.2]{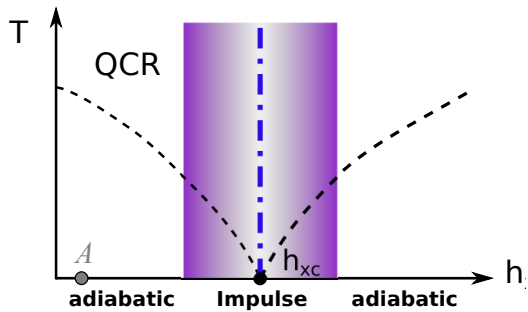}
   \caption{\label{drivenregion}(Color online) Comparison of different initial conditions. Point A (grey), which is far away from the critical point $h_{\rm xc}$, signals the initial state for the original quantum KZM. The dash-dotted line (blue), which is exactly at the critical point and parallel with the vertical axis, is the starting state in the driven dynamics studied in Ref.~\onlinecite{Polkov1,Polkov2,Dengss}. In our work, the starting positions can be distributed over the whole impulse region and its finite-temperature extension (purple).}
\end{figure}

Developing effective theories to describe the non-equilibrium phenomena in quantum systems is of central significance in condensed matter physics and ultracold atom physics~\cite{revqkz1,revqkz2,duttabook}. For instance, the Kibble-Zurek mechanism (KZM), which was originally proposed by Kibble in cosmology~\cite{Kibble1}, and then by Zurek in condensed matter physics~\cite{Zurek1}, has been generalized to describe the driven quantum critical dynamics starting with a state far from the critical point~\cite{qkz1,qkz2,qkz3,qkz4,qkz5,qkz6,qkz7,qkz8,qkz9,Chandran,qkz10,Zhong1,Zhong2}. The driven critical dynamics are controlled by the competition between the energy gap and  the external driving. While the energy gap tends to suppress the excitation from the ground state, the external driving tends to create the excitation. Accordingly, the KZM separates the whole driven process into two adiabatic stages and an impulse stage, as sketched in Fig.~\ref{drivenregion}. Near the quantum critical point, the gap becomes very small and the external driving dominates. This results in an impulse stage, in which macroscopical excitations are created by the external driving. In contrast, far away from the quantum critical point, the energy gap is large enough to suppress the excitation and the system evolves adiabatically. This results in an adiabatic stage. The KZM demonstrates that the number of topological defects, which are generated during the impulse stage, scales with the driving rate~\cite{revqkz1,revqkz2,duttabook}. Recently, the KZM has been verified experimentally in trapped-ion systems~\cite{Ulm,Pyka} and Bose-Einstein condensates~\cite{Navon,Clark}. The full scaling forms has been employed to numerically detect the critical properties in both classical~\cite{Zhong1,Zhong2,CWLiu1,CWLiu2} and quantum phase transitions~\cite{qkz10,Hu}.

For the original KZM, it is necessary to have an initial adiabatic stage before an impulse stage to have a frozen correlation length that generates topological defects~\cite{revqkz1,revqkz2,duttabook}. To go beyond KZM, it is important to investigate, both theoretically and experimentally, to what extent the scaling theory should be modified when the initial adiabaticity breaks down. For classical phase transitions, effects induced by a non-equilibrium initial state near the critical point have been studied~\cite{Huang,Feng}. However, the conclusion cannot be directly generalized to the real-time quantum critical dynamics, because the dynamical properties are intrinsically different in quantum and classical cases. For quantum phase transitions, the driven critical dynamics starting exactly from a quantum critical point has been investigated~\cite{Polkov1,Polkov2,Dengss,revqkz1,revqkz2}. In reality, however, such a scenario is the exception rather than the norm. Studies on driven quantum critical dynamics starting from the vicinity of a quantum critical point are hence called for.

Another important issue is the thermal effects in quantum critical dynamics~\cite{Sachdevbook}. A scaling theory that includes the temperature has been proposed in Ref.~{\onlinecite{Chandran}}, in which the dimensional analysis is used to derive the scaling form but effects induced by the physical procedure to bring the temperature into the dynamics is not considered. However, it has been demonstrated that how the thermal effects are taken into account can affect the dynamic scaling form\cite{Yin,Patane}. In general, there are two natural setups to include the thermal effects. One approach is to consider an open quantum system in which the quantum critical system is coupled to an infinite heat bath of temperature $T$. Within this approach, it has been demonstrated that the scaling theory must include the dissipation rate as a scaling variable~\cite{Yin,Yin2}. The scaling theory is hence different from the theory in Ref.~{\onlinecite{Chandran}}. The other approach is to consider the critical dynamics in a closed quantum system starting from a thermal equilibrium state. This approach has been used in Refs.~\onlinecite{Polkov1,Polkov2,Dengss}. However, in these studies the initial state is always the thermal equilibrium state exactly at the quantum critical point. Furthermore, the scaling behavior during the whole driven process has not been discussed therein. It is thus imperative to investigate whether the scaling theory proposed in Ref.~{\onlinecite{Chandran}} can characterise more general cases of driven dynamics in closed systems.

In this work, we consider the quantum critical dynamics of a closed system under an external driving parameterized $g=g_0+R_gt$. Here $g$ is the distance of the relevant parameter to the critical point, $g_0$ is the initial value of $g$ and is near the critical point, and $R_g$ is the driving rate. Both zero temperature and finite temperature initial states are considered, as shown in Fig.~\ref{drivenregion}. By imposing a scale transformation on the master equation, we propose a general scaling theory. Comparing with the original quantum KZM, we find that this scaling theory includes $g_0$ and the initial temperature $T$ as additional scaling variables. There are two important difference between the present scaling theory and the scaling relation proposed in Ref.~\onlinecite{Polkov1,Polkov2,Dengss}. First, the present scaling theory generalizes the initial condition to the whole scaling regime which is controlled by the fixed point corresponding to the critical point. In particular, the initial state can be in the impulse region, as depicted in Fig.~\ref{drivenregion}. Second, it shows that the scaling behavior exists during the whole driven process. We also point out that $g_0$ plays a crucial role in the driven dynamics when the thermal effects are considered. For small $g_0$, the scaling behavior for an equilibrium initial state with a finite temperature $T$ exists and can be described by our present theory. For very large $g_0$, the initial temperature $T$ is irrelevant, and the scaling can be described by the original KZM. In between these two limits, however, the driven dynamics cannot be described by simple scaling forms. These results fill the vacancies of the scaling theory proposed in Ref.~{\onlinecite{Chandran}}. We confirm our scaling theory numerically by taking the one-dimensional ($1$D) transverse-field Ising model as an example.

The rest of the paper is organized as follows. In Sec.~\ref{sctheory}, we propose the scaling theory of driven critical dynamics starting in the vicinity of the critical point and compare it with the original quantum KZM. Then, in Sec.~\ref{ver}, we numerically verify the scaling theory by taking $1$D transverse-field Ising model as an example. We compare our scaling theory with the previous studies in Sec.~\ref{Dis}. Finally, a summary is given in Sec.~\ref{sum}.

%%%%%
\section{\label{sctheory}Scaling theory}
We consider the driven dynamics in a closed quantum system starting with a thermal equilibrium state at temperature $T$ near the quantum critical point. This setup has been considered in Ref.~\onlinecite{Polkov1,Polkov2,Dengss,Polkov3,Cardy}. Recently the rapid advancement of the experimental technologies has enabled us to isolate well a quantum system from the environment~\cite{Newton}.  The scenario studied here is hence within the reach of current experiments. We start from the master equation that governs the evolution of the density matrix of the system:

\begin{equation}
\frac{\partial \rho(t)}{\partial t}=-i\left[H(g(t)), \rho(t)\right]. \label{mastereq}
\end{equation}
We assume that the systems starts from the equilibrium thermal state associated with the initial Hamiltonian. The initial density matrix is
\begin{equation}
\rho(0)=\frac{{\rm exp}(-H(g_0)/T)}{{\rm Tr}({\rm exp}(-H(g_0)/T))}, \label{inicon}
\end{equation}
which is the density matrix of the ground state when temperature is zero. The evolution of the expectation value for a operator $Y$ at any temperature is calculated by
\begin{equation}
\left\langle Y\right\rangle(t)={\rm Tr}\{Y \rho[\rho(0),g(R_g,g_0,t)]\}. \label{expval}
\end{equation}

Since the initial position is near the critical point and the driving rate is small~\cite{Silvi}, the driven process is expected to demonstrate scaling behaviors. To explore scaling properties of the driven critical dynamics, we impose a scale transformation, with a rescaling factor $b$ on Eq.~(\ref{mastereq}). Under this scale transformation~\cite{Sachdevbook}, $t\rightarrow \tilde{t}=tb^{-z}$, $H\rightarrow \tilde{H}=Hb^{z}$, $g\rightarrow \tilde{g}=gb^{1/\nu}$, $R_g\rightarrow \tilde{R}_g=R_gb^{r_g}$~\cite{Yin}, $T\rightarrow \tilde{T}=Tb^{z}$, and $\rho$ is unchanged as it is dimensionless. From the dimension of $g$ and $t$, one obtains $r_g=z+1/\nu$. By replacing the variables in Eq.~(\ref{mastereq}) with the corresponding rescaled ones, we obtain the rescaled master equation
\begin{equation}
\frac{\partial \rho(\tilde{t})}{\partial \tilde{t}}=-i\left[\tilde{H}(\tilde{g}(\tilde{t})), \rho(\tilde{t})\right], \label{remastereq}
\end{equation}
with the rescaled initial condition
\begin{equation}
\rho(0)=\frac{{\rm exp}(-\tilde{H}(\tilde{g}_0)/\tilde{T})}{{\rm Tr}({\rm exp}(-\tilde{H}(\tilde{g}_0)/\tilde{T}))}. \label{reinicon}
\end{equation}
Meanwhile, the rescaled operator $\tilde{Y}$, defined as $\tilde{Y}\equiv Yb^s$, should satisfy
\begin{equation}
\langle \tilde{Y}\rangle(\tilde{t})=b^s{\rm Tr}\{Y\rho[\rho(0),\tilde{g}(\tilde{R}_g,\tilde{g}_0,\tilde{t})]\}, \label{rescalY}
\end{equation}
according to Eq.~(\ref{remastereq}). Comparing Eq.~(\ref{expval}) with Eq.~(\ref{rescalY}), we find that the scale transformation of $\left\langle Y\right\rangle$ under external driving $g=g_0+R_gt$ reads
\begin{eqnarray}
\begin{aligned}
&\left\langle Y\right\rangle(g,g_0,R_g,T,t)=\\&b^{-s}\left\langle \tilde{Y}\right\rangle(gb^{1/\nu},g_0b^{1/\nu},R_gb^{r_g},Tb^z,tb^{-z}). \label{scaY}
\end{aligned}
\end{eqnarray}

By comparing with the KZM~\cite{qkz6,qkz7,qkz8,qkz9,Chandran,qkz10} one finds that when $T=0$ and $|g_0|\gg R_g^{-1/\nu r}$ the scale transformation~(\ref{scaY}) falls back to the the scaling theory of the original quantum KZM. The initial stage is an adiabatic stage with correlation length $|g_0|^{-\nu}$ and at $\hat{g}\equiv R_g^{1/\nu r_g}$ the evolution crosses over to an impulse region with correlation length $R_g^{-1/r_g}$. In this limit the initial $g_0$ is not a necessary scaling variable. This is because in the initial adiabatic stage the role played by $g_0$ can be replaced by $g$ and the effects induced by $R_g$ can be ignored.

In general, however, there are two non-trivial effects described by Eq.~(\ref{scaY}). Firstly, Eq.~(\ref{scaY}) is applicable when $g_0$ is in the impulse region, i.e., $|g_0|\ll R_g^{-1/\nu r_g}$. In contrast to the driven dynamics for large $|g_0|$, here $g_0$ is an indispensable scaling variable. This is because the initial gap $|g_0|^{\nu z}$, which impedes the excitation induced by the external driving, is smaller than the driven energy $R_g^{z/r_g}$, which promotes the excitations. As a result, there are jumping from ground state to excited states since the beginning. Moreover, the intensity of this initial jump is determined by the competition between $|g_0|^{\nu z}$ and $R_g^{z/r_g}$. Therefore, both $g_0$ and $R_g$ will affect the driven dynamics.

Secondly, the initial temperature $T$ has been included in Eq.~(\ref{scaY}) as an additional scaling variable. We emphasize that for $T\neq0$ the non-trivial scaling behavior only exists when $g_0$ is in the vicinity of the quantum critical point. To clarify this cooperative effect between $g_0$ and $T$, we compare the thermal effects in this case and in other two cases with different $g_0$.  \textit{Case A:} For very large $|g_0|$ and non-zero $T$, the universal behavior is similar to the original KZM in which the driven dynamics starts with the ground state for large $|g_0|$. The reason is that for very large $|g_0|$, the initial gap is also very large and the initial thermal excitation can be neglected. Therefore, the following evolution is almost identical to the evolution starting with the ground state due to the unitary property of the dynamics. Accordingly, for large $|g_0|$, the initial temperature is irrelevant and the scaling theory falls back to the original KZM. \textit{Case B:} When $|g_0|$ is chosen to be a medium value, the thermal effects are of significance but they do not posses any scaling behavior that is determined by the critical point. The dynamics is quite complicated, because the scale transformation in Eq.~(\ref{reinicon}) is not applicable any more. As a consequence the driven dynamics cannot be described by Eq.~(\ref{scaY}). In contrast, when $g_0$ is near the critical point as considered above, $T$ plays a significant role since the gap is small. Furthermore, the scale transformation imposed on the initial condition, i.e., Eq.~(\ref{reinicon}), is applicable and the dynamics shows scaling behaviors. To sum up we tabulate the results in Table.~\ref{tab1}, from which one finds that non-trivial driven scaling behavior with an initial thermal state only exists when $g_0$ is small. This cooperative effect between the roles played by $g_0$ and $T$ has not been discussed in Ref.~{\onlinecite{Chandran}} and our scaling theory makes up this missing link.

\begin{table}[htbp]
  \centering
  \caption{Driven dynamics from thermal equilibrium states with different $|g_0|$.}
    \begin{tabular}{c l| c | c }
    \hline
    \hline
    \multicolumn{2}{c|}{$|g_0|$} & scaling behavior & thermal effects  \\
    \hline
          & very large     & $\surd$   &   $\times$  \\
          & medium     & $\times$  &  $\surd$ \\
          & very small    & $\surd$  &  $\surd$  \\
    \hline
    \end{tabular}%
  \label{tab1}%
\end{table}%

To be explicit, we consider the scaling behavior of the evolution of the order parameter $M$. According to Eq.~(\ref{scaY}), for $g=g_0+R_gt$ with $g_0$ being near the critical point, the scale transformation of $M$ is
\begin{eqnarray}
\begin{aligned}
&M(t,R_g,g_0,g,T)=\\&b^{-\beta/\nu}M(tb^{-z},R_gb^{r_g},g_0 b^{1/\nu},g b^{1/\nu},Tb^z), \label{scalet}
\end{aligned}
\end{eqnarray}
where $b$ is a rescaling factor. By assuming $R_gb^{r_g}=1$, we obtain the scaling form of the order parameter,
\begin{eqnarray}
\begin{aligned}
&M(R_g,g_0,g,T)=\\&R_g^{\beta/\nu r_g}f_1(g_0 R_g^{-1/\nu r_g},g R_g^{-1/\nu r_g},TR_g^{-z/r_g}), \label{scalingf}
\end{aligned}
\end{eqnarray}
where $f_i$ ($i=1$ for the present case) is the scaling function.

For small $g_0$ at zero temperature the driven dynamics of the von-Neumann entanglement entropy also demonstrates a scaling behavior affected by the initial conditions. The von-Neumann entanglement entropy is measured as $S=-\textrm{Tr}(\rho\textrm{log}\rho)$, where $\rho$ is the reduced density matrix of half of the system. For a $1$D system near its quantum critical point it has been show that the entanglement entropy scales as $S=(c/6)\textrm{log}\xi$~\cite{Eisert,Amico,Osterloh,Laorencie,Cala}, where $c$ is the central charge and $\xi$ is the correlation length.
%Recently, the entanglement entropy has been measured in experiments~\cite{Islam}. {\bf not sure if we want to talk about the experiment, do they measure entanglement or Renyi entropy?}

For $g=g_0+Rt$, according to Eq.~(\ref{scaY}), we can write down the scale transformation of the correlation length,
\begin{equation}
\xi(t,g,g_0,R)=b\xi(tb^{-z},gb^{1/\nu},g_0b^{1/\nu},R_gb^{r_g}). \label{xisc}
\end{equation}
When $R_gb^{r_g}=1$, i.e., $b=R_g^{-1/r_g}$, Eq.~(\ref{xisc}) gives the scaling form of $\xi$ under the external driving,
\begin{equation}
\xi(g,g_0,R)=R_g^{-1/r_g}f_2(gR_g^{-1/\nu r_g},g_0R_g^{-1/\nu r_g}). \label{xisf}
\end{equation}
Therefore, the entanglement entropy $S$ satisfies
\begin{eqnarray}
\begin{aligned}
&S(g,g_0,R_g)=\\&-\frac{c}{6r_g}\textrm{log}R_g+f_3(gR_g^{-1/\nu r_g},g_0R_g^{-1/\nu r_g}). \label{eesc}
\end{aligned}
\end{eqnarray}

\section{\label{ver} Numerical verification of the scaling theory}
In this section, we numerically verify the scaling theory proposed for the universal driven dynamics starting in the vicinity of the critical point.

\subsection{\label{monum}Model and numerical method}
To illustrate our scaling theory, in the following, we will take the $1$D transverse field Ising model as an example. The Hamiltonian reads~\cite{Sachdevbook}
\begin{equation}
H_I=-\sum\limits_{n}\sigma_n^z\sigma_{n+1}^z-h_x\sum\limits_{n}\sigma_n ^x,
\label{HIsing}
\end{equation}
where $\sigma_n^x$ and $\sigma_n^z$ are the Pauli matrices in $x$ and $z$ direction, respectively,
at site $n$, $h_x$ is the transverse field. We have set the
Ising coupling to unity as our energy unit. The order parameter is defined
as $M=\langle\sigma^z_n\rangle$, where the angle brackets denote the average of the operator over
each site. The critical point of model~(\ref{HIsing}) is
$h_{xc}=1$. The distance to the critical point $g$ is $g\equiv h_x-h_{xc}$. The exact critical exponents are $\beta=1/8$, $\nu=1$, $z=1$~\cite{Sachdevbook}, $r_g=2$, and the central charge $c=1/2$~\cite{Eisert,Amico}. This model has been realized in CoNb$_2$O$_6$ experimentally~\cite{Coldea}.

The infinite time-evolving block decimation (iTEBD) algorithm~\cite{Vidali} is used to calculate the evolutions of the order parameter and the entanglement entropy for the zero temperature situation. According to this algorithm, a quantum state is represented by a matrix product state via Vidal's decomposition. In this way, each site is attached by a matrix. The evolution of a state then is realized by the updating of these matrices according to the local evolution operators, which are obtained by the Suzuki-Trotter decomposition of $\exp(-iHt)$.

For the driven dynamics starting from a thermal equilibrium state, we need to purify the identity matrix $\mathcal{I}$ into a pure state $|\phi(0)\rangle_\mathcal{I}$ by introducing an auxiliary system. The state in this auxiliary system should be maximally entangled to the physical system. In this way, by tracing the freedom in the auxiliary system, the density matrix recovers the identity matrix $\mathcal{I}$. It has been proved that the expectation value of a operator $Y$ in a thermal equilibrium state at temperature $T$ can be calculated via~\cite{Cirac,Scho}
\begin{equation}
\left\langle Y\right\rangle_T=\frac{_\mathcal{I}\langle\phi(T)| Y|\phi(T)\rangle_\mathcal{I}}{_\mathcal{I}\langle\phi(T)|\phi(T)\rangle_\mathcal{I}},
\label{thermaleq}
\end{equation}
where $|\phi(T)\rangle_\mathcal{I}\equiv {\rm exp}(-H/2T)|\phi(0)\rangle_\mathcal{I}$. In the same way, we can calculate the real-time evolution starting from a thermal equilibrium state as
\begin{equation}
\left\langle Y\right\rangle_T(t)=\frac{_\mathcal{I}\langle\phi(T,t)| Y|\phi(T,t)\rangle_\mathcal{I}}{_\mathcal{I}\langle\phi(T,t)|\phi(T,t)\rangle_\mathcal{I}},
\label{thermaleq2}
\end{equation}
where $|\phi(T,t)\rangle_\mathcal{I}\equiv \mathcal{T}{\rm exp}(-iH(t)t)|\phi(T)\rangle_\mathcal{I}$, in which $\mathcal{T}$ is the time-ordering operator.

In the following calculations, the time interval is chosen as $0.005$ and $100$ states are kept for both zero temperature and finite temperature cases.

\subsection{\label{nr}Numerical results}
\begin{figure}
  \centerline{\epsfig{file=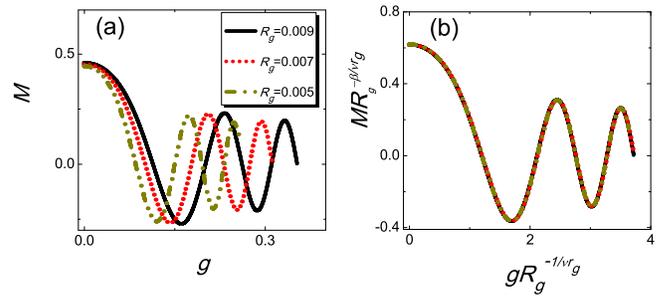,width=1.0\columnwidth}}
  \caption{\label{changeot} (Color online) The evolution of $M$ under increasing $g$ with fixed $g_0R_g^{-1/\nu r_g}=-0.01054$ for three $R_g$ indicated. The curves before and after rescaled are shown in (a) and (b) respectively.}
\end{figure}

\begin{figure}
  \centerline{\epsfig{file=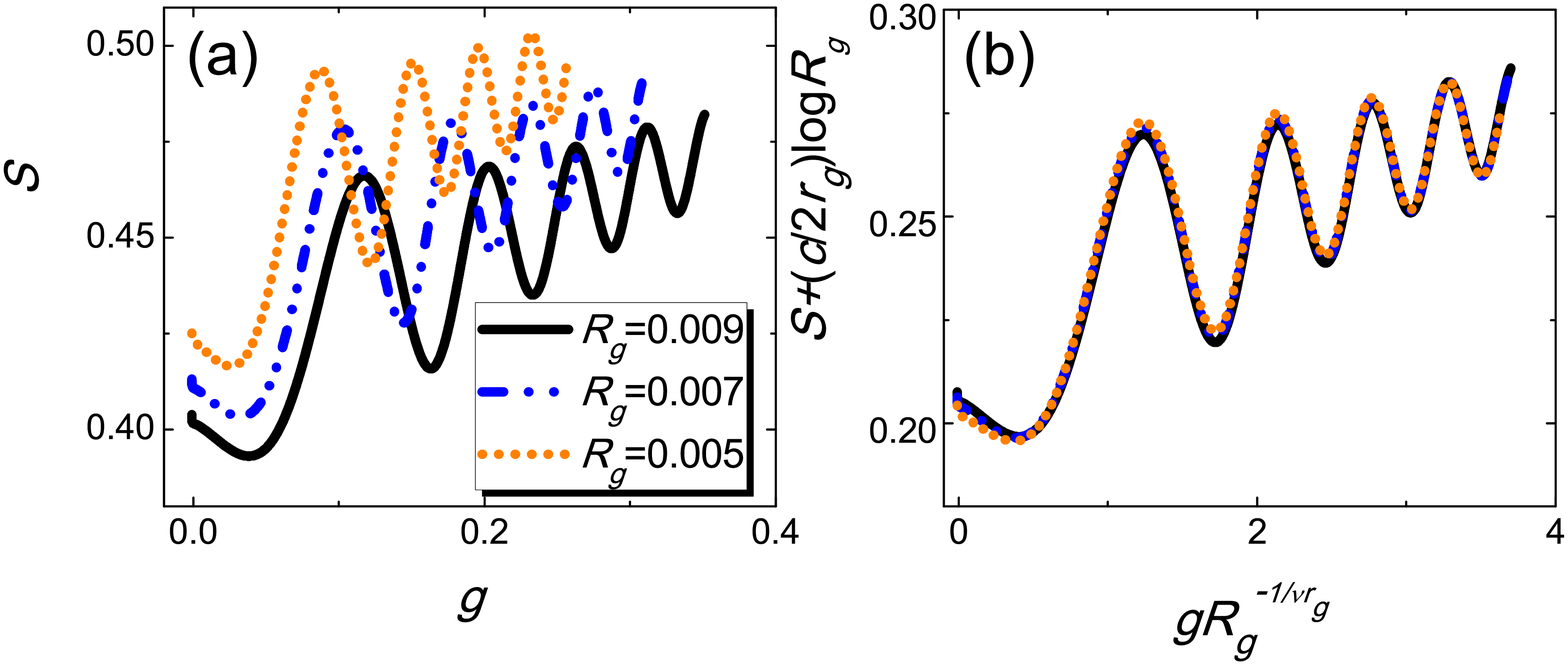,width=1.0\columnwidth}}
  \caption{\label{changegotee} (Color online) The evolution of $S$ under increasing $g$ with fixed $g_0R_g^{-1/\nu r_g}=-0.01054$ for three $R_g$ indicated is shown in (a). The curves for $S+(c/6r_g)\textrm{log}R_g$ versus the rescaled $g$ are shown in (b).}
\end{figure}

Figure~\ref{changeot} shows curves of $M$ versus $g$ for different driving rates with fixed $g_0R_g^{-1/\nu r_g}$ at $T=0$. These curves collapse onto each other after rescaling according to Eq.~(\ref{scalingf}). This demonstrates that the scaling behavior exists although the evolution curves are quite different from the original KZM. The scaling theory is also confirmed in Fig.~\ref{changegotee}, in which we show that curves of $S+(c/6r_g)\textrm{log}R_g$ versus $gR_g^{-1/\nu r_g}$ with fixed $g_0R_g^{-1/\nu r_g}$ match with each other, confirming Eq.~(\ref{eesc}).

\begin{figure}
  \centerline{\epsfig{file=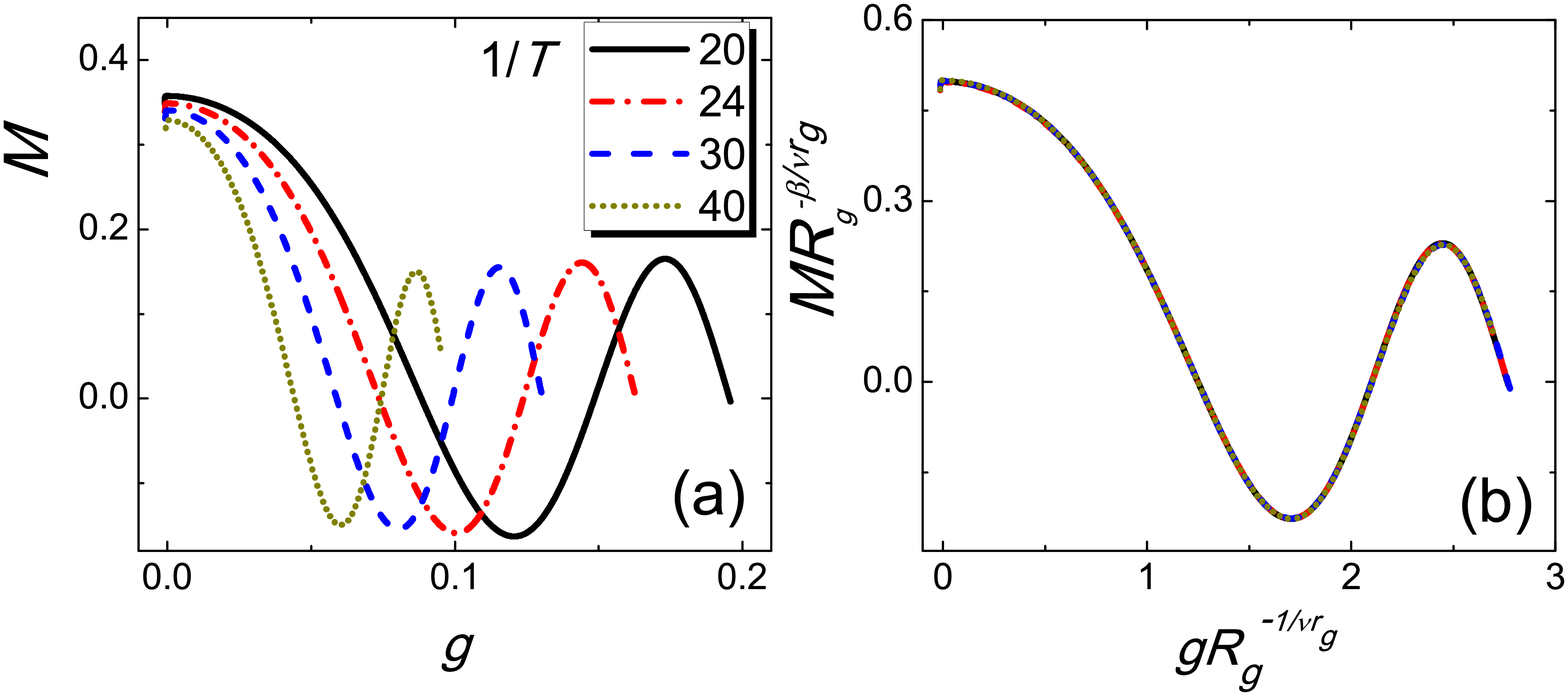,width=1.0\columnwidth}}
  \caption{\label{changegft} (Color online) The evolution of $M$ under increasing $g$ with fixed $g_0R_g^{-1/\nu r_g}=0.0141$, $h_0R_g^{-\beta\delta/\nu r_g}=-0.0718$ and $T^{-1}R_g^{z/r_g}=0.7071$ for four $T$ indicated. The curves before and after rescaled are shown in (a) and (b) respectively.}
\end{figure}

Figure~\ref{changegft} examines the scaling form of Eq.~(\ref{scalingf}) including $T$. In model~(\ref{HIsing}) the magnetization vanishes for any finite temperature according to the Mermin-Wagner theorem. So we have imposed a symmetry breaking field $h_0$ on the system. For fixed $g_0R_g^{-1/\nu r_g}$, $h_0R_g^{-\beta\delta/\nu r_g}$ ($\delta=15$ \cite{Sachdevbook}) and $T^{-1}R_g^{z/r_g}$, curves at different temperatures perfectly overlap with each other after rescaling, confirming Eq.~(\ref{scalingf}) including the initial temperature $T$.

\begin{figure}
  \centerline{\epsfig{file=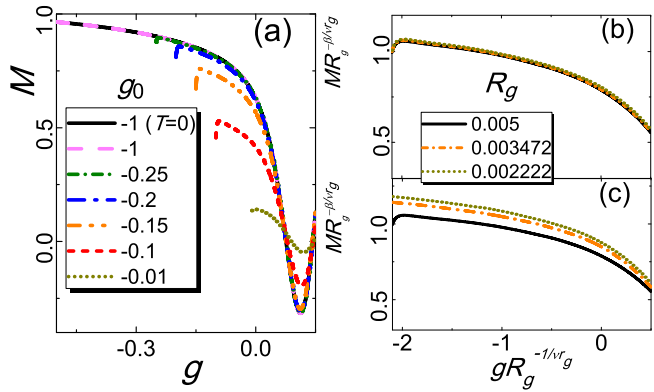,width=1.0\columnwidth}}
  \caption{\label{crossover} (Color online) (a) $M$ versus $g$ for different $g_0$ with $h_0=0.0005$, $R_g=0.005$, and $1/T=10$ (except the black solid curve with $1/T=\infty$ as marked). (b) $MR_g^{-\beta/\nu r_g}$ versus $gR_g^{-1/\nu r_g}$ with fixed $g_0R_g^{-1/\nu r_g}=-2.121$ (when $g_0=-0.15$, $R_g=0.005$.), $h_0R_g^{-\beta\delta/\nu r_g}=0.07181$, and $T^{-1}R_g^{z/\nu r_r}=0.3535$. (c) $MR_g^{-\beta/\nu r_g}$ versus $gR_g^{-1/\nu r_g}$ with fixed $g_0=-0.15$, $h_0R_g^{-\beta\delta/\nu r_g}=0.07181$, and $T^{-1}R_g^{z/\nu r_r}=0.3535$.}
\end{figure}

Figure~\ref{crossover} shows crossover effects of the temperature $T$ from the large $|g_0|$ to small $|g_0|$. From Fig.~\ref{crossover}(a), one finds that for $h_0=0.0005$, $R_g=0.005$ and $1/T=10$, when $g_0<-0.25$, the curves match with the curve for $1/T=\infty$ and very large $|g_0|$. Thus, $T$ is irrelevant for large $|g_0|$. When $g_0>-0.2$, the thermal effects play significant roles. In particular, for $ -0.15 < g_0 <0$, as displayed in Fig.~\ref{crossover}(b), the critical dynamics is controlled by the critical point and the initial temperature should be rescaled according to Eq.~(\ref{scalingf}). The narrow region, $-0.2<g_0<-0.15$, is the crossover region, in which the dynamics cannot be described by simple power laws. Furthermore, we show in Fig.~\ref{crossover}(c) that when $g_0$ is not rescaled, curves of $MR_g^{-\beta/\nu r_g}$ versus $gR_g^{-1/\nu r_g}$ cannot collapse, even when $g=0$. This indicates that $g_0$ is an indispensable scaling variable when $T$ contributes significantly to the evolution. This supplements the scaling theory discussed in Ref.~\onlinecite{Chandran}.

\section{\label{Dis}Discussion}
We compare our scaling theory with the reported results in Refs.~\onlinecite{Polkov1,Polkov2,Dengss}. First, Refs.~\onlinecite{Polkov1,Polkov2,Dengss} have studied the driven critical dynamics beginning with a thermal equilibrium state exactly at the quantum critical point. In contrast, our theory is applicable for any starting position near the critical point. Second, Refs.~\onlinecite{Polkov1,Polkov2,Dengss} studied the scaling behavior of the topological defects after the quench, however, in the present paper, our scaling theory describes the scaling behavior in the whole process. To quantitatively compare our theory with the previous ones, we consider the number of topological defects for $g=g_0+R_gt$ with $g_0=0$. According to Eq.~(\ref{scaY}), the scaling form for the number of topological defects is
\begin{equation}
n(R_g,g,T)=R_g^{d/r_g}f_4(g R_g^{-1/\nu r_g},TR_g^{-z/r_g}). \label{scalingdef}
\end{equation}
At the impulse-adiabatic boundary $\hat{g}\equiv R_g^{-1/\nu r_g}$, $n(R_g,\hat{g},T)=R_g^{d/r_g}f_5(TR_g^{-z/r_g})$. Comparing with the results in Refs.~\onlinecite{Polkov1,Polkov2,Dengss}, we obtain the form of $f_5(TR_g^{-z/r_g})$ for different kinds of excitations,
\begin{equation}
 \label{excitation}
  f_5(x) = \left \{
\begin{array}{l l}
x,  &  {\rm bosonic}\\
& \\
x^{-1},  & {\rm fermionic}
\end{array}
\right.
\end{equation}

Another setup to generalize the quantum KZM to the finite-temperature region is considering an open quantum system in which the system is attached to an infinite heat bath with a temperature $T$~\cite{Yin}. Comparing these two scaling theories, we find the following differences. First, the dissipation rate is an indispensable scaling variable in the driven criticality of the open quantum system~\cite{Yin}, while there is no dissipative effect when the finite-temperature is only an initial condition as we considered here. Second, in the open quantum system~\cite{Yin}, the initial stage can be far away from the critical point, while in the present case, the initial thermal equilibrium state is near the critical point. In both cases, the critical behavior is controlled by the original critical point of the Hamiltonian, so there is no other additional critical exponent introduced.

\section{\label{sum}Summary}
We have studied the driven critical dynamics starting near the quantum critical point. A scaling theory is developed to describe the scaling behavior in the whole driving process. In this scaling theory, the initial relevant variables are included in the scaling form as indispensable scaling variables. Besides the case at zero temperature, we have also explored the driven dynamics starting with a thermal equilibrium state near the critical point and found that the scaling behavior can be described by the scaling theory in which the initial temperature is rescaled according to its equilibrium scale transformation. We have verified our scaling theory by taking the $1$D transverse-field Ising model as an example. Since real experiments is implemented at finite temperatures, our scaling theory can be experimentally examined. For some experiments in which the usual KZM cannot work, our scaling theory also provides a possible candidate to modify the scaling analysis.

\section{\label{Ack}Acknowledgments}
We acknowledge the support by Ministry of Science and Technology (MOST) of Taiwan through Grant No. 104-2628-M-007-005-MY3. We also acknowledge the support from the National Center for Theoretical Science (NCTS) of Taiwan.


\begin{thebibliography}{99}
\bibitem{revqkz1} J. Dziarmaga, Adv. Phys. {\bf 59}, 1063 (2010).
\bibitem{revqkz2} A. Polkovnikov, K. Sengupta, A. Silva, and M. Vengalattore, Rev. Mod. Phys. {\bf 83}, 863 (2011).
\bibitem{duttabook} A. Dutta, G. Aeppli, B. K. Chakrabarti, U. Divakaran, T. F. Rosenbaum, and D. Sen {\it Quantum Phase Transitions in Transverse Field Spin Models: From Statistical Physics to Quantum Information}, (Cambridge University Press, 2015).
\bibitem{Kibble1} T. W. B. Kibble, J. Phys. A: Math. Gen. {\bf 9}, 1387 (1976).
\bibitem{Zurek1} W. H. Zurek, Nature {\bf 317}, 505 (1985).
\bibitem{qkz1} W. H. Zurek, U. Dorner, and P. Zoller, Phys. Rev. Lett. {\bf 95}, 105701 (2005).
\bibitem{qkz2} J. Dziarmaga, Phys. Rev. Lett. {\bf 95}, 245701 (2005).
\bibitem{qkz3} A. Polkovnikov, Phys. Rev. B {\bf 72}, 161201(R) (2005).
\bibitem{qkz4} B. Damski and W. H. Zurek, Phys. Rev. Lett. {\bf 99}, 130402 (2007).
\bibitem{qkz5} D. Sen, K. Sengupta, and S. Mondal, Phys. Rev. Lett. {\bf 101}, 016806 (2008).
\bibitem{qkz6} S. Deng, G. Ortiz, and L. Viola, Europhys. Lett. {\bf 84}, 67008 (2008).
\bibitem{qkz7} C. De Grandi, A. Polkovnikov, and A. W. Sandvik, Phys. Rev. B {\bf 84}, 224303 (2011).
\bibitem{qkz8} M. Kolodrubetz, B. K. Clark, and D. A. Huse, Phys. Rev. Lett. {\bf 109}, 015701 (2012).
\bibitem{qkz9} M. Kolodrubetz, D. Pekker, B. K. Clark, and K. Sengupta, Phys. Rev. B {\bf 85}, 100505(R) (2012).
\bibitem{Chandran} A. Chandran, A. Erez, S. S. Gubser, and S. L. Sondhi, Phys. Rev. B {\bf 86}, 064304 (2012).
\bibitem{qkz10} S. Yin, X. Qin, C. Lee, and F. Zhong, arXiv:1207.1602.
\bibitem{Zhong1} S. Gong, F. Zhong, X. Huang, and S. Fan, New J. Phys. {\bf 12}, 043036 (2010).
\bibitem{Zhong2}  F. Zhong, in {\it Applications of Monte Carlo Method in Science and Engineering}, Edited by S. Mordechai (InTech, Rijeka, 2011).
\bibitem{Ulm} S. Ulm, J. Ro{\ss}nagel,	G. Jacob, C. Deg\"{u}nther,	S. Dawkins, U. Poschinger, R. Nigmatullin, A. Retzker, M. Plenio, F. Schmidt-Kaler, and K. Singer,	Nat. Commun. {\bf 4}, 2290 (2013).
\bibitem{Pyka} K. Pyka, J. Keller, H. L. Partner, R. Nigmatullin, T. Burgermeister, D. M. Meier, K. Kuhlmann, A. Retzker, M. B. Plenio, W. H. Zurek, A. del Campo, and T. E. Mehlst\"{a}ubler, Nat. Commun. {\bf 4}, 2291 (2013).
\bibitem{Navon}N. Navon, A. L. Gaunt, R. P. Smith, and Z. Hadzibabic, Science {\bf 347}, 167 (2015).
\bibitem{Clark}L. W. Clark, L. Feng, C. Chin, arXiv: 1605.01023.
\bibitem{CWLiu1} C. Liu, A. Polkovnikov, and A. Sandvik, Phys. Rev. Lett. {\bf 114}, 147203 (2015).
\bibitem{CWLiu2} C. Liu, A. Polkovnikov, A. Sandvik, and A. Young, Phys. Rev. E {\bf 92}, 022128 (2015).
\bibitem{Hu} Q. Hu, S. Yin, and F. Zhong, Phys. Rev. B {\bf 91}, 184109 (2015).
\bibitem{Huang} Y. Huang, S. Yin, Q. Hu, and F. Zhong, Phys. Rev. B {\bf 93}, 024103 (2016).
\bibitem{Feng} B. Feng, S. Yin, and F. Zhong, arXiv:1604.04345.
\bibitem{Polkov1} C. De Grandi, V. Gritsev, and A. Polkovnikov, Phys. Rev. B {\bf 81}, 012303 (2010).
\bibitem{Polkov2} C. De Grandi, V. Gritsev, and A. Polkovnikov,  Phys. Rev. B {\bf 81}, 224301 (2010).
\bibitem{Dengss} S. Deng, G. Ortiz, and L. Viola, Phys. Rev. B {\bf 83}, 094304 (2011).
\bibitem{Sachdevbook} S. Sachdev, {\it Quantum Phase Transitions}, (Cambridge University Press, 1999).
\bibitem{Yin} S. Yin, P. Mai, and F. Zhong, Phys. Rev. B, {\bf 89}, 094108 (2014).
\bibitem{Patane}D. Patan\`{e}, A. Silva, L. Amico, R. Fazio, and G. E. Santoro, Phys. Rev. Lett. {\bf 101}, 175701 (2008).
\bibitem{Yin2} S. Yin, C.-Y. Lo, and P. Chen, Phys. Rev. B {\bf 93}, 184301 (2016)
\bibitem{Polkov3} A. Polkovnikov and V. Gritsev, Nat. Phys. 4, 477 (2008).
\bibitem{Cardy} S. Sotiriadis, P. Calabrese, and J. Cardy, Europhys. Lett. {\bf 87}, 20002 (2009).
\bibitem{Newton}T. Kinoshita, T. Wenger, and D. S. Weiss, Nature {\bf 440}, 900 (2006).
\bibitem{Silvi} P. Silvi, G. Morigi, T. Calarco, and S. Montangero, Phys. Rev. Lett. {\bf 116}, 225701 (2016).
\bibitem{Eisert} J. Eisert, M. Cramer and M. Plenio, Rev. Mod. Phys. {\bf 82}, 277 (2010).
\bibitem{Amico} L. Amico, R. Fazio, A. Osterloh, and V. Vedral, Rev. Mod. Phys. {\bf 80}, 517 (2008).
\bibitem{Osterloh} A. Osterloh, L. Amico, G. Falci, and R. Fazio, Nature {\bf 416}, 608 (2002).
\bibitem{Laorencie} N. Laorencie, arXiv:1512.03388.
\bibitem{Cala}P. Calabrese and J. Cardy, J. Stat. Mech. P04010 (2005).
%\bibitem{Islam}R. Islam, R. Ma, P. M. Preiss, M. Eric Tai, A. Lukin, M. Rispoli, and M. Greiner, Nature {\bf 528}, 77 (2015).
\bibitem{Coldea} R. Coldea, D. A. Tennant, E. M. Wheeler, E. Wawrzynska, D. Prabhakaran, M. Telling, K. Habicht, P. Smeibidl, and K. Kiefer, Science {\bf 327}, 177 (2010).
\bibitem{Vidali} G. Vidal, Phys. Rev. Lett. {\bf 98}, 070201 (2007).
\bibitem{Cirac}  F. Verstraete, J. Garc\'{\i}a-Ripoll, J. Cirac, Phys. Rev. Lett. {\bf 93}, 207204 (2004).
\bibitem{Scho} U. Schollw\"{o}ck, Annals of Physics, {\bf 326}, 96 (2011).
\end{thebibliography}
\end{document}